\documentclass{mem}
\usepackage{natbib}\usepackage{txfonts}\usepackage{balance}
\usepackage{graphicx}
\usepackage[a4paper,breaklinks,dvipdfm]{hyperref}
\idline{75}{282}
\begin{document}

\title{Lithium abundances in extremely metal-poor turn-off stars}


\author{
L. \, Sbordone\inst{1,2} 
P. \, Bonifacio\inst{2}
and E. \, Caffau\inst{1,2}
          }


\institute{
Landessternwarte Heidelberg, K\"onigstuhl 12, 69117 Heidelberg, Germany
\and
GEPI, Observatoire de Paris, CNRS, Univ. Paris Diderot, Place
Jules Janssen, 92190
Meudon, France\\
\email{lsbordon@lsw.uni-heidelberg.de}
}

\authorrunning{Sbordone, Bonifacio, Caffau}

\titlerunning{Li in EMP TO stars}

\abstract{
We discuss the current status of the sample of Lithium abundances in extremely metal poor (EMP) turn-off (TO) stars collected by our group, 
and compare it with the available literature results. In the last years, evidences have accumulated of a progressive disruption of the Spite plateau
in stars of extremely low metallicity. What appears to be a flat, thin plateau above [Fe/H]$\sim$-2.8 turns, at lower metallicities, into a 
broader distribution for which the plateau level constitutes the upper limit, but more and more stars show lower Li abundances. The sample we have
collected currently counts abundances or upper limits for 44 EMP TO stars between [Fe/H]=-2.5 and -3.5, plus the ultra-metal poor star 
SDSS J102915+172927 at [Fe/H]=-4.9. The ``meltdown'' of the Spite plateau is quite evident and, at the current status of the sample, does not appear
to be restricted to the cool end of the effective temperature distribution. SDSS J102915+172927 displays an extreme Li depletion that contrasts with its 
otherwise quite ordinary set of [X/Fe] ratios. 
\keywords{be nice: write down your keywords here}
}
\maketitle{}

\section{Introduction}

Since its first discovery by \citet{spite82}, the so-called Spite plateau has constituted a puzzling observational fact. The very fact that Li could be measurable in old, warm stars was unexpected, since atomic diffusion was presumed to deplete it in the atmosphere to the level of non-detectability \citep{michaud84}. The long-standing interpretation of the Li abundance plateau in old stars as being representative of the primordial Big Bang Nucleosynthesis (BBN) yield was further challenged by the measurements issued of the WMAP satellite \citep{iocco09}. Its measurement of the photon-baryon ratio at the time of the BBN leads to estimate a primordial Li abundance about 0.5 dex higher than observed. A promising explanation of said discrepancy is the adoption of a theory of ``mitigated''  atomic diffusion in stellar atmospheres, where turbulent mixing reduces the efficiency of Li settling \citep[e. g.][and references therein]{lind09}. However, current models of said effect are still entirely parametric, limited to a restricted parameter space, and do not reproduce satisfactorily the variations of Li abundance with the evolutionary state of the stars \citep[see][]{gonzalez09}.  

\begin{figure*}[t!]
\resizebox{\hsize}{!}{\includegraphics[clip=true]{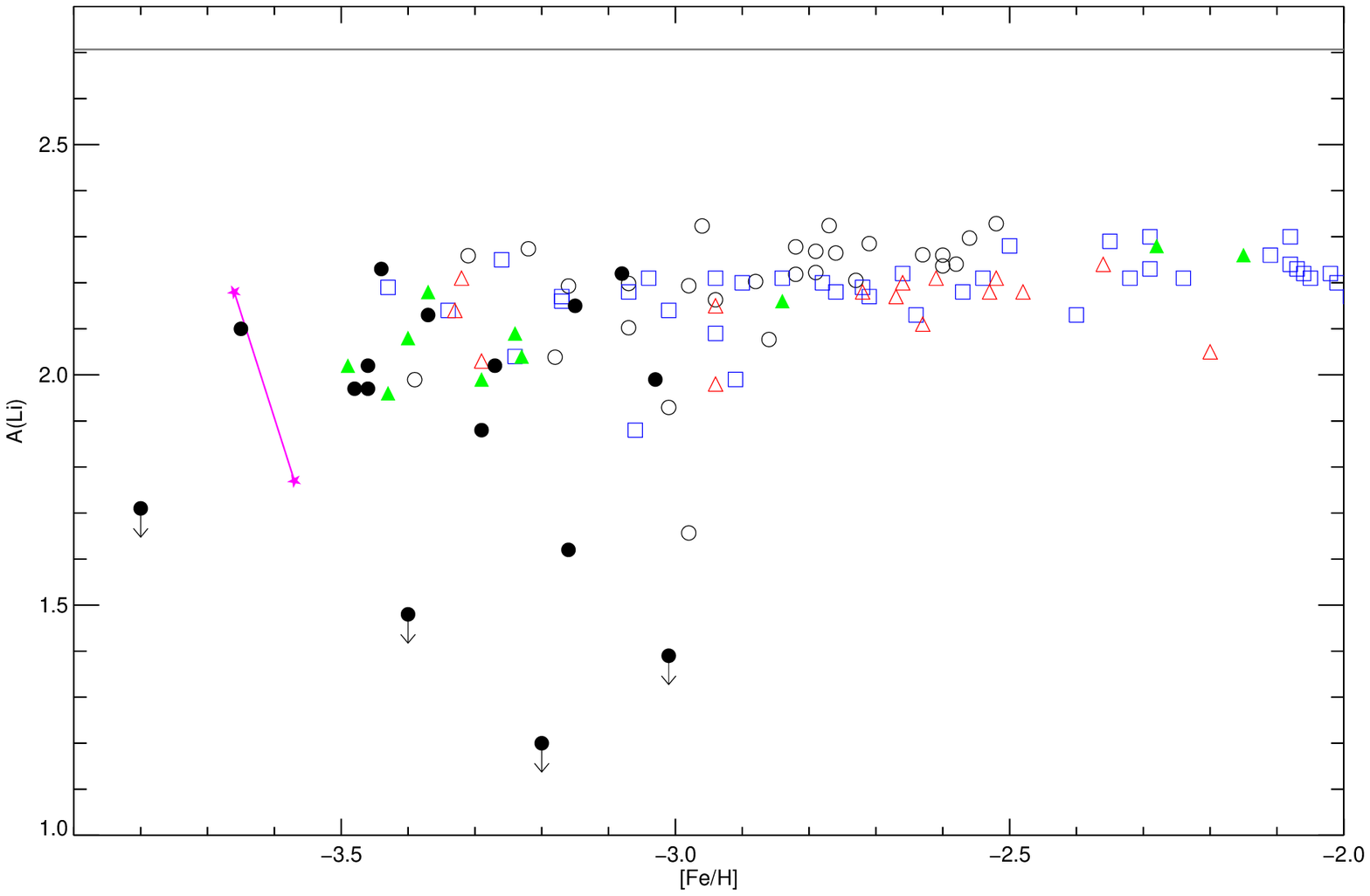}}
\caption{\footnotesize
A(Li) vs. [Fe/H] for various samples. Filled black circles, \citet{bonifacio12}; open black circles{sbordone10}; green filled triangles \citet{aoki09}; red open triangles \citet{hosford09}; magenta stars, the two components of the binary star CS 22876-032 \citep{gonzalez08}, blue open squares \citet{asplund06}. Points with a downward arrow indicate upper limits. The grey horizontal line indicates the current estimated primordial Li abundance based on WMAP results.}
\label{fig1}
\end{figure*}

At the same time, evidences have been accumulating that the Spite plateau is disrupted at extremely low metallicities. In \citet{sbordone10} we showed how warm (5800 K$<$T$_{\mathrm{eff}}$$<$6800 K) TO stars at metallicities roughly below [Fe/H]=-2.8 do no longer show the uniform Li abundance typical of the plateau, but rather scatter along an increasingly dispersed distribution for which the plateau abundance appears to constitute the upper limit. Previous claims that dispersion in Li abundance increased for EMP, or that the plateau appeared tilted \citep[e.g.][]{asplund06,aoki09}, appeared now as low-statistics manifestations of what we called the ``meltdown'' of the Spite plateau (see Fig. \ref{fig1}). In a different take to the same problem, \citet{melendez10} noticed how more metal poor stars of the same mass and evolutionary status have higher temperatures, and interpreted the observed effects at the EMP end as due to mass-dependent diffusive depletion. To account for this, \citet{melendez10} proposed to exclude from samples used to evaluate the Spite plateau any star for which T$_{\mathrm{eff}}$$<$5850-180*[Fe/H]. By using a sample ``cleaned'' accordingly to said cutoff, \citet{melendez10} recovered a thin plateau, unaffected by the ``meltdown'', but broken in two at [Fe/H]$\sim$-2.5, the metal rich section showing a Li abundance about 0.05 dex higher than the metal poor one. However, \citet{melendez10} sample below [Fe/H]=-3 was much more restricted than the one analyzed in \citet{sbordone10}.

\begin{figure*}[t!]
\resizebox{\hsize}{!}{\includegraphics[clip=true]{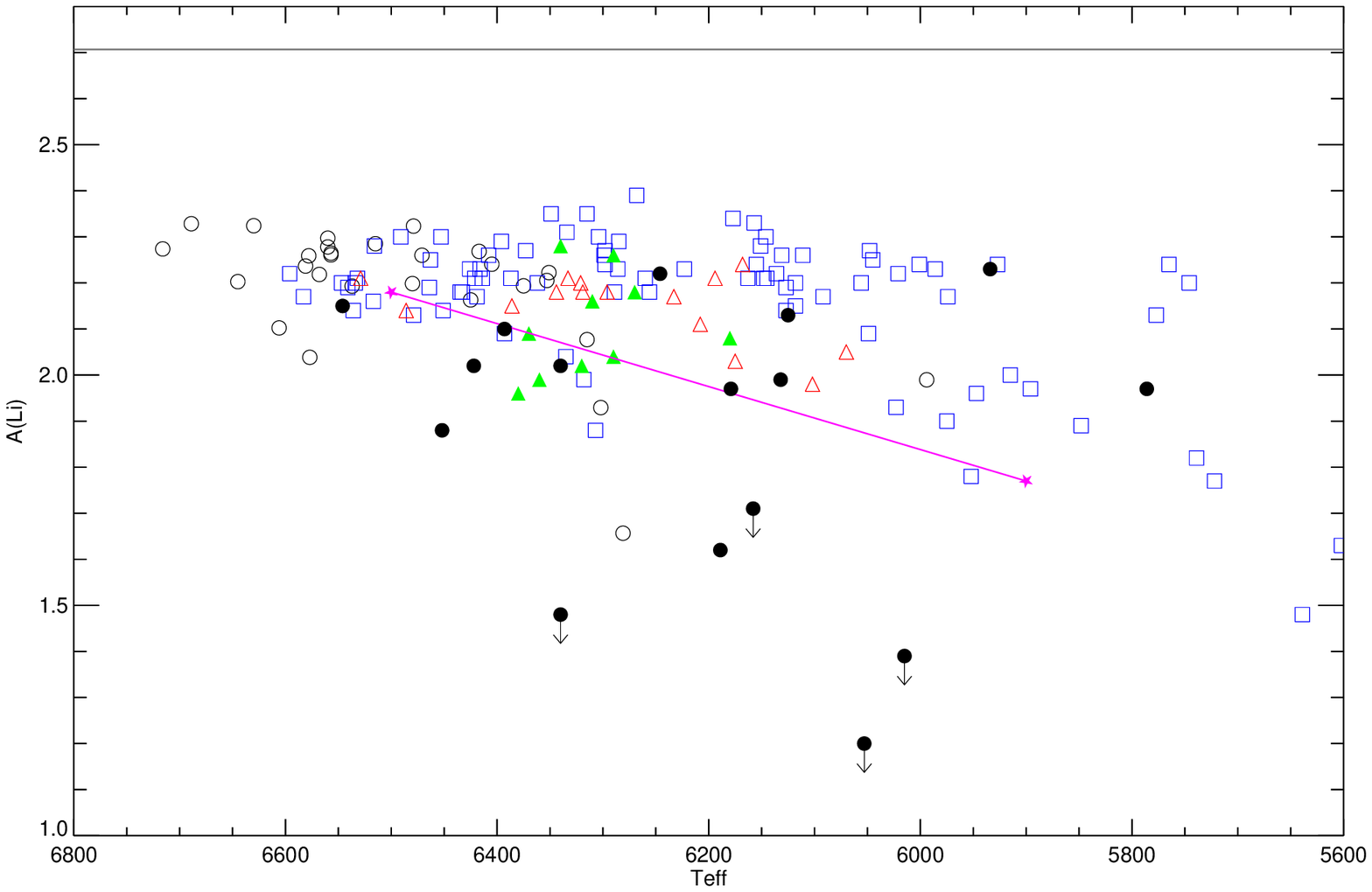}}
\caption{\footnotesize
A(Li) vs. T$_{\mathrm{eff}}$ for various samples. Symbols as in Fig. \ref{fig1}}
\label{fig2}
\end{figure*}

\section{The current sample}

Since some years, we have undertaken an effort to identify extremely metal poor stars from low resolution Sloan Digital Sky Survey (SDSS) spectra, and follow them up with high resolution spectroscopy \citep{ludwig08,bonifacio12}. A number of papers have been issued of the so-far analyzed sample: \citet{behara10} analyzed 3 carbon enhanced metal poor stars, \citet{caffau11a} and \citet{bonifacio11} presented the sub-sample observed with X-SHOOTER, \citet{caffau11b,caffau12} described the ultra metal poor (UMP) star SDSS J102915+172927, and \citet{bonifacio12} presented the largest sub-sample, 16 stars observed with UVES. The present work describes Li abundances for stars in \citet{bonifacio12} and SDSS J102915+172927. 

The stars in \citet{bonifacio12} have been analyzed by means of the automatic abundance analysis code MyGIsFOS \citep{sbordone12} to derive atmospheric parameters and chemical abundances. Lithium abundances have been derived thanks to the 3D NLTE Li 670 nm doublet fitting function presented in \citet{sbordone10}, which has been employed also for estimating the Li abundance upper limit in SDSS J102915+172927. In both cases, effective temperature has been derived from (g-z)$_{0}$ colors, gravity from Fe I -- Fe II ionization equilibrium, when measurable Fe II lines were present, or set to $\log $g=4 otherwise.

\section{Results}

By including the new sample together with the \citet{sbordone10} one, the evidence of the Spite plateau ``meltdown'' below [Fe/H]$\sim$-2.8 is further reinforced (Fig. \ref{fig1}). The \citet{sbordone10} sample supported to some extent the \citet{melendez10} claim that, at extremely low metallicities, progressively hotter stars should be expected to show Li abundances below the plateau level: all the stars in the sample below T$_{\mathrm{eff}}$=6000 K, in fact, showed a Li abundance below the plateau. The situation, however was not a clear cut one, since also a number of relatively hot stars appeared depleted. In the \citet{boifacio12} sample, on the other hand, at least one star with T$_{\mathrm{eff}}$$<$6000 K lies on the Spite plateau (Fig. \ref{fig2}). 

\begin{figure*}[t!]
\resizebox{\hsize}{!}{\includegraphics[clip=true]{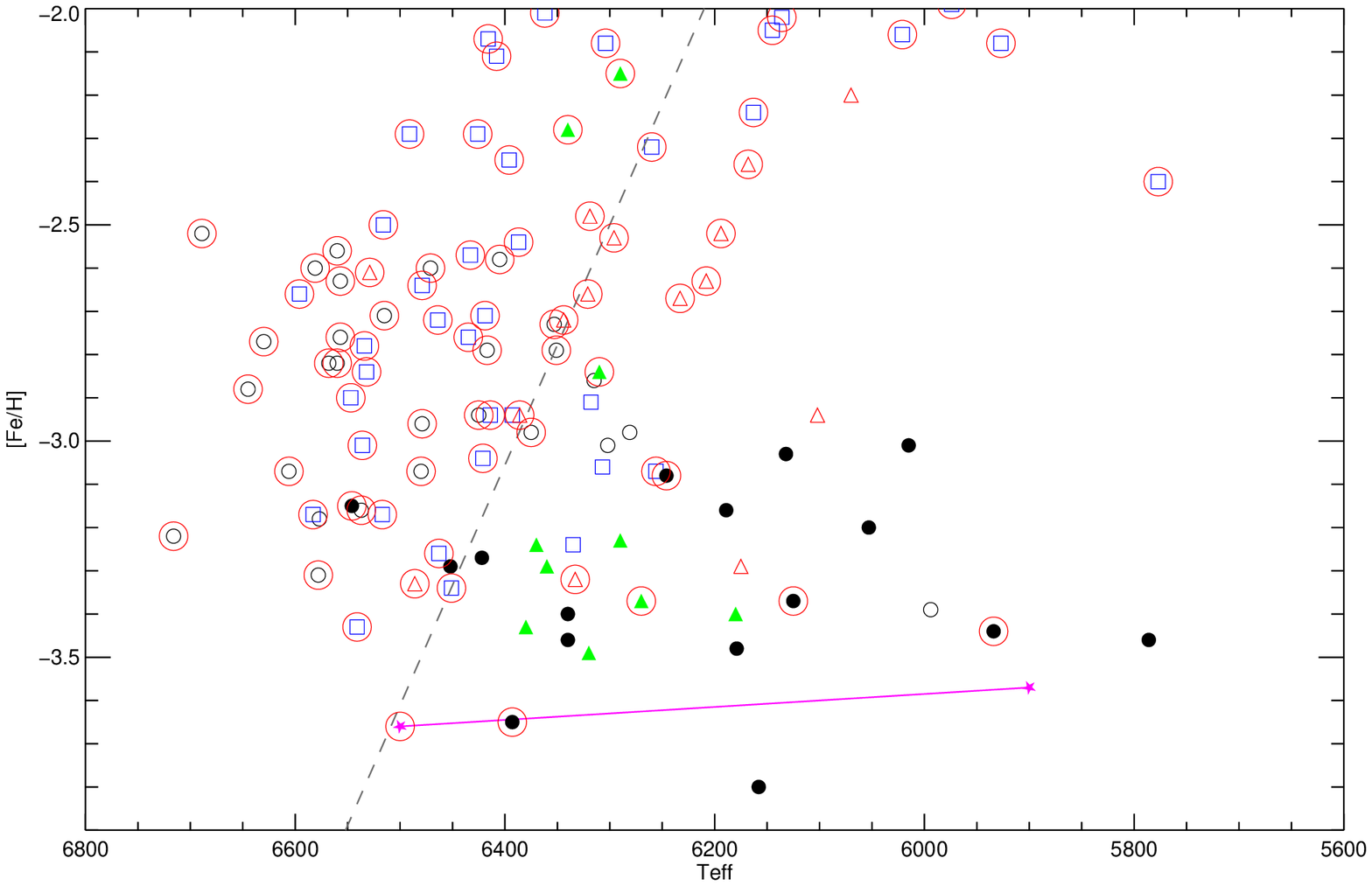}}
\caption{\footnotesize
[Fe/H] vs. T$_{\mathrm{eff}}$ for the stars in Fig. \ref{fig1} and \ref{fig2}. Symbols as in Fig. \ref{fig1}. The grey dashed line represents the discriminant proposed by \citet{melendez10}, i.e. T$_{\mathrm{eff}}$=5850-180*[Fe/H]. The symbols encyrcled in red correspond to stars with A(Li)$>$2.1.}
\label{fig3}
\end{figure*}

\begin{figure*}[t!]
\resizebox{\hsize}{!}{\includegraphics[clip=true]{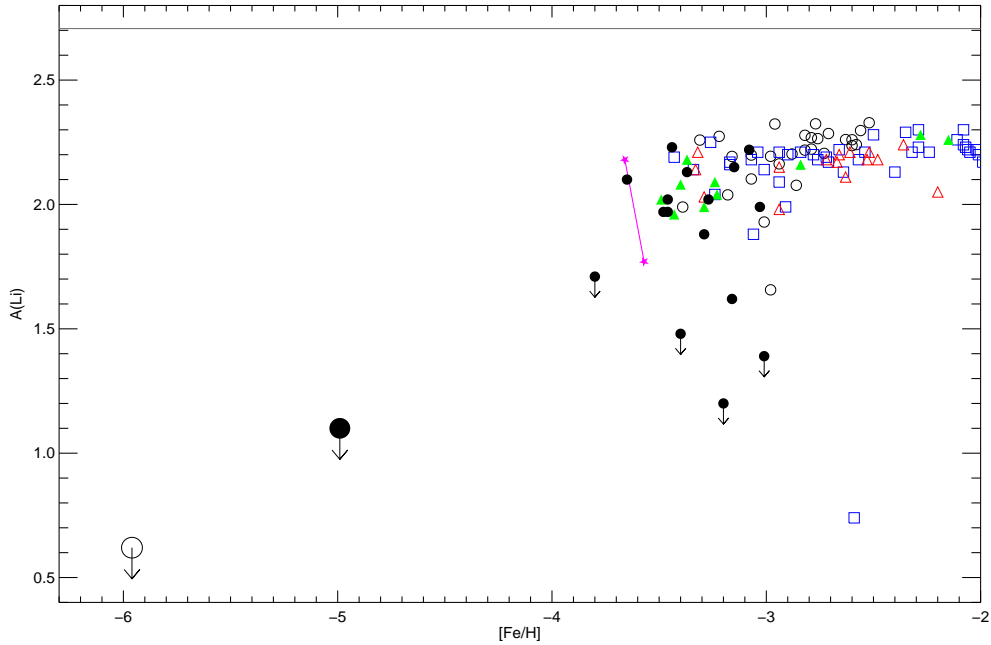}}
\caption{\footnotesize
Like in Fig. \ref{fig1} but now including the two ultra metal poor stars HE 1327-2326 and SDSS J102915+172927 (open and filled large circles, respectively). Symbols for the other stars are the same as in Fig \ref{fig1}.}
\label{fig4}
\end{figure*}

To try to clarify the issue, we plot in Fig. \ref{fig3} the effective temperature versus  [Fe/H] for the stars in the various samples already presented in Fig. \ref{fig1} and \ref{fig2}. The grey dashed line represents the discriminant proposed by \citet{melendez10}, a red circle surrounds all the stars whose Li abundance exceeds 2.1 dex, i.e. falling on the plateau. In the considered samples, about 42\% pf the stars falling below the \citet{melendez10} discri1327minant fall, in fact, on the Spite plateau. The \citet{melendez10} discriminant appears to fail, in particular, above [Fe/H]=-2.8, where almost all the stars it cuts out fall on the Spite plateau: we consider this one further hint that some remarkably different physical process affects the Li abundance in stars above and below said metallicity. But even beolow [Fe/H]=-2.8, still about one third of the sampled stars falling below the \citet{melendez10} discriminant appear to fall on the plateau, indication that whatever mechanism is causing the meltdown, it is not affecting ll the stars uniformly as a function of their temperature and metallicity alone. 

In Fig. \ref{fig4} we reproduce Fig. \ref{fig1}, but adding the two ultra metal poor stars known to date for having atmospheric parameters compatible with the Spite plateau, HE 1327-2326 \citep{frebel08} and SDSS J102915+172927 \citep{caffau11b,caffau12}. In both these stars the Li 670nm doublet is not detected, and only an upper limit is derived. Both these upper limits, moreover, correspond to values much lower than the Spite plateau. One might, in principle, consider discounting HE 1327-2326 since the star presents an extreme CNO enhancement, indication of significant pollution by CNO-cycled material in the nebula out of which the star formed. Due to the high temperature characteristic of the CNO cycle, one expects such material to be essentially Li-free. SDSS J102915+172927, on the other hand, presents no detectable CNO enhancement, and its abundance ratios are the ones of a normal $\alpha$-enhanced metal poor stars, except for the particularly low metallicity. One is thus entitle to consider said star as an indicator that the trend towards increasing Li depletion continues in the [Fe/H]$<$-4 range.

\section{Conclusions}

The ``meltdown'' of the Spite plateau is confirmed by the expanded sample of EMP To stars presented in \citet{bonifacio12}, as well as by the discovery of SDSS J102915+172927. Its origins remail unclear. Mechanisms invoking atmospheric effects, such as diffusion, to explain it, face a formidable challenge: decreasing the metallicity, metallic opacities in atmospheres, as well as the effect of metals as electron donors, become increasing negligible. Below [Fe/H]=-2.5, a change in metallicity leaves the atmospheric structure largely unchanged, while the same is not true at higher metallicities. And yet, the Spite plateau appears to exist between [Fe/H]=-2 and -1, but is disrupted below [Fe/H]=-2.8. The structure of SDSS J102915+172927 should closely resemble the one of a comparable [Fe/H]=-3 star, while the same is not true between [Fe/H]=-2.5 and [Fe/H]=-1. Nevertheless, Li abundance in SDSS J102915+172927 is much lower than in a typical [FeH]=-3 star of similar parameters, while a thin plateau exist at higher metallicities. It is apparent to us that some piece of the puzzle is still missing. It is possible that the missing ingredient is related to the formation environment of extremely and ultra metal poor stars. It is currently believed that the more metallicity decreases, the more the initial mass function should become top-heavy. It is thus possible that the formation of low mass stars requires increasingly particular conditions at lower metallicities. One might for example speculate that protostellar nebula fragmentation might be favored by high level of turbulence, leading to low mass stars being preferentially formed as very fast rotators, thus increasing Li depletion due to higher convective layer depth. Further investigation in the physics of extremely low metallicity star formation could possibly lead to clarify this aspect.



\end{document}